\documentclass[onecolumn,showpacs,showkeys,preprintnumbers,amsmath,amssymb]{revtex4}
\usepackage{graphicx}
\usepackage{dcolumn}
\usepackage{bm}
\usepackage{epsfig}
\usepackage{rotating}

\begin{document}
\title{Hierarchical structure of the European countries based on debts as a percentage of GDP during the 2000-2011 period}

\author{Ersin Kantar$^{a, b}$, Bayram Deviren$^{c}$, and Mustafa Keskin$^{*, b}$
\\
$^{a}$Institute of Science, Erciyes University, 38039 Kayseri, Turkey
\\
$^{b}$Department of Physics, Erciyes University, 38039 Kayseri, Turkey
\\
$^{c}$Department of Physics, Nev\c{s}ehir University, 50300
Nev\c{s}ehir, Turkey
}

\altaffiliation[]{Corresponding Author: Department of Physics, Erciyes University, 38039 Kayseri, Turkey
\\Email: keskin@erciyes.edu.tr (M. Keskin);
\\Tel.: +90 352 4374938x33105;
\\Fax: +90352 4374931.}

\begin{abstract}We investigate hierarchical structures of the European countries by using debt as a percentage of Gross Domestic Product (GDP) of the countries as they change over a certain period of time. We obtain the topological properties among the countries based on debt as a percentage of GDP of European countries over the period 2000-2011 by using the concept of hierarchical structure methods (minimal spanning tree, (MST) and hierarchical tree, (HT)). This period is also divided into two sub-periods related to 2004 enlargement of the European Union, namely 2000-2004 and 2005-2011, in order to test various time-window and observe the temporal evolution. The bootstrap techniques is applied to see a value of statistical reliability of the links of the MSTs and HTs. The clustering linkage procedure is also used to observe the cluster structure more clearly. From the structural topologies of these trees, we identify different clusters of countries according to their level of debts and economic ties. Our results show that by the debt crisis, the less and most affected Eurozone's economies are formed as a cluster with each other in the MSTs and hierarchical trees.
\end{abstract}

\maketitle

\section{Introduction}
Financial integration has increased dramatically over the past decade, especially among advanced economies. There has also been an increasing presence of foreign intermediaries in several banking systems (including many emerging markets). The interconnection in the global financial system means that if one nation defaults on its sovereign debt or enters into recession thus putting some external private debt at risk, the banking systems of creditor nations face losses. For example, in October 2011 Italian borrowers owed French banks \$366 billion (net). Should Italy be unable to finance itself, the French banking system and economy could come under significant pressure, which in turn would affect France's creditors and so on. This is referred to as financial contagion \cite{Seth,Alan}. As a result, international risk sharing and efficiency among the economies of countries have increased. From late 2009, fears of a sovereign debt crisis developed among fiscally conservative investors concerning some European states, with the situation becoming particularly tense in early 2010 \cite{Haidar,Matlock}. This included Eurozone members Greece, Ireland and Portugal and also some European Union (EU) countries outside the area. Iceland, the country which experienced the largest crisis in 2008 when its entire international banking system collapsed, has emerged less affected by the sovereign debt crisis as the government was unable to bail the banks out. In the EU, especially in countries where sovereign debt has increased sharply due to bank bailouts, a crisis of confidence has emerged with the widening of bond yield spreads and risk insurance on credit default swaps between these countries and other EU members.

On the other hand, complex networks have been able to successfully describe the topological properties and characteristics of many real-life systems \cite{Ray2007,Garas2008,Wang2009,Kantar2011}. Moreover, there are studies which focus on understanding the complex structure of stock markets and financially growing systems; these topology \cite{Bonanno2003}, viscoelastic behavior \cite{Gündüz2010} anomalous diffusion \cite{Meerschaert2006}, phase plots \cite{Chiarella2006}, phase transition \cite{Serrano2007}, wavelet techniques \cite{Caetano2007}, the Le Chatelier principle \cite{Lady2007}, non-equilibrium dynamics \cite{Özlale2007}, networks \cite{Albert2002}, graph theory \cite{Aste2006}, quantum field theory and path integrals \cite{Baaquie2006,Decamps2006}, uncertainty \cite{Schinckus2009}, and spin models \cite{Yang2006}. The properties of economic crisis have been previously studied by using correlation networks \cite{Naylor2007,Schiavo2010,Yu2010,Lisewski2010,Aste2010}. However,
to the best of our knowledge, use of the countries, network to analyze an economic crisis has only been examined in a few studies, such as \cite{Dias2011,Lee2011}. Dias \cite{Dias2011} analyzed the topology of correlation networks among countries based on daily yield rates on 10-year government bonds for nineteen EU using the concept of an MST and HT for the 2007-2010 period and three sub-period, namely 2007-2008, 2008-2009 and full year of 2010. He performed a technique to associate the value of statistical reliability to the links of the MSTs and HTs by using bootstrap technique. Moreover, Lee \textit{et al.} \cite{Lee2011} investigated the crisis spreading dynamics by using the GDP and the international trade data of the countries in the 2002-2006 period.

On the other hand, hierarchial structure of the European countries based on debts as a percentage of GDP during the 2000-2011 period is not investigated. Therefore, the purpose of this work is to investigate hierarchical structures of the European countries by using debt as a percentage of Gross Domestic Product (GDP) of the countries as they change over a certain period of time. We study three time-window data sets, the periods of 2000-2004 and 2005-2011 and 2000-2011 and observe the temporal evolution of the European debt crisis by using the concept of the minimal spanning tree (MST) and hierarchical tree (HT). The reason for studying sub-periods is due to the enlargement of the European Union in 2004. The geometrical and taxonomic information about the
correlation between the elements of the set can be obtained from the MST and HT, respectively. These methods were successfully applied to analyze
currency, equity and commodity markets. The notion of distance introduced in these applications is based on the Pearson correlation coefficient as a function to measure the similarity between two time series. We also used the bootstrap technique to quantify the statistical reliability of hierarchical trees. Finally, we applied average linkage cluster analysis (ALCA) to clearly observe the clusters of countries.
The MST, HT and ALCA  give a useful guide to define the underlying economic
or regional causal connections for individual countries.

Mantegna \cite{Mantegna1999}, and
Mantegna and Stanley \cite{Mantegna2000} introduced the MST and HT that have been  previously used to
analyze currency markets
\cite{Mizuno2006,Brida2009a,Feng2010,Keskin2010a},
in particular to find the clustered structure of currencies and the key
currency in each cluster \cite{Mizuno2006,Naylor2007,Feng2007,
Keskin2010a} and to resolve contagion in a currency crisis
\cite{Brida2009a}. These trees are also applied to investigate the
clustering behavior of individual stocks within a single country
\cite{Bonanno2004,Brida2010,Brida2010c}.
The MST and the HT has also been used to study world equity markets
\cite{Bonanno2000,Coelho2007a}, European equity markets
\cite{Gilmore2008} and commodity markets
\cite{Sieczka2009,Tabak2010}. Finally, the dynamic MST analysis has also been developed and applied
to investigate the time-varying behavior of stocks.
\cite{Onnela2003a,Onnela2003b,Onnela2002,Micciche2003}.

In correlation based hierarchical investigations, the bootstrap approach has been used
to quantify the statistical reliability of hierarchical trees
and correlation based networks. Laloux \textit{et al.} \cite{Laloux1999} and Plerou \textit{et al.} \cite{Plerou1999}
applied Random Matrix Theory methods to obtain quantitative
estimation of the statistical uncertainty of the correlation matrix.
Moreover, the bootstrap approach was used to
quantify the statistical reliability of hierarchical trees and
correlation based networks by Tumminello \textit{et al.}
\cite{Tumminello2007b,Tumminello2007a,Tumminello2010}. In addition, Tumminello \textit{et al.} \cite{Tumminello2011} investigated the statistical assessment of links in bipartite complex systems. The topology of correlation networks among 34 major currencies using the concept of an MST and HT, and bootstrap replicas of data studied by Keskin \textit{et al.} \cite{Keskin2010a}.  Moreover, Kantar \textit{et al.} applied the bootstrap technique to investigate the topological properties of Turkey's foreign trade \cite{Kantar2011} as well as the major international and Turkish companies \cite{Kantar2011a}.

Finally, correlation based clustering has been used to infer the hierarchical structure of a
portfolio of stocks from its correlation coefficient matrix \cite{Mantegna1999,Bonanno2003,Bonanno2001}.
Useful examples of correlation based networks apart from the
minimal spanning tree are the planar maximally filtered graph
\cite{Tumminello2005} and the average linkage minimal spanning tree
\cite{Tumminello2007a,Kantar2011,Kantar2011a}.

The outline of the remaining part of this paper is organized as follows. Section II
introduces the methodology and the sampling procedures while Section
III shows the data and Section IV presents empirical results. Finally,
Section V provides some final considerations.

\section{Methodology}
We describe the basics of construction of the MST, HT and ALCA. First, we generate a network of countries by using the MST and the HT that is obtained starting from the MST. Second, we examined the statistical reliability and stability of our results by using the bootstrap technique. Finally, we investigated cluster structures within the average linkage cluster analysis.

\subsection{Minimal spanning tree and hierarchical trees}

Since the construction of a minimal spanning tree (MST) and hierarchical tree (HT), which is also called single linkage cluster analysis (SLCA) has been described extensively in Mantegna and Stanley \cite{Mantegna2000} as well as in our previous papers \cite{Keskin2010a,Kantar2011,Kantar2011a}, therefore we shall only give a brief summary here.

The correlation function between a pair of countries based on the debts of European countries in order to quantify synchronization between the countries is defined as

\begin{equation} \label{GrindEQ__1_}
C_{ij} =\frac{\left\langle R_{i} R_{j} \right\rangle -\left\langle R_{i} \right\rangle \left\langle R_{j} \right\rangle }{\sqrt{\left(\left\langle R_{i}^{2} \right. \rangle -\left\langle R_{i} \right\rangle ^{2} \right)\left(\left\langle R_{j}^{2} \right. \rangle -\left\langle R_{j} \right\rangle ^{2} \right)} } ,
\end{equation}
\noindent
where i and j are the numerical labels to the debts of countries and the notation $\left\langle ...\right\rangle $ means an average over time. ${ R}_{{ i}}$ is the vector of the time series of log-returns, ${ R}_{{ i}} {(t)\; =\; ln\; P}_{{ i}} { (t\; +\; }\tau { )\; -\; ln\; P}_{{ i}} { (t)}$ is the log-return and ${ P}_{{ i}} (t)$ is the quarterly debt ratio of country i at quarter t. All cross-correlations range from -1 to 1, where -1 and +1 mean that two countries i and j are completely anti-correlated and correlated, respectively. In the case of $C_{ij} $ = 0 the countries i and j are uncorrelated.

The MST is based on the idea that the correlation coefficient between a pair of countries can be transformed to a distance between them by using an appropriate function as a metric. An appropriate function for this transformation is

\begin{equation} \label{GrindEQ__2_}
{\rm d}_{{\rm ij}} =\sqrt{2(1-C_{ij} )} ,
\end{equation}
\noindent
where ${\rm d}_{{\rm ij}} $ is a distance for a pair of the rate i and the rate \textit{j}. Now, one can construct an MST for a pair of countries using the N x N matrix of ${\rm d}_{{\rm ij}} $.

On the other hand, to construct an HT, we used the ultrametric distance, which introduced by Mantegna \cite{Mantegna2000}, or the maximal ${\rm d}_{{\rm ij}}^{{\rm \wedge }} $${}_{ }$ between two successive countries encountered when moving from the first country \textit{i} to the last country \textit{j} over the shortest part of the MST connecting the two countries. The distance fulfills the condition ${\rm d}_{{\rm ij}}^{{\rm \wedge }} {\rm \; }\le {\rm max\{ d}_{{\rm ik}} {\rm ,\; d}_{{\rm kj}} {\rm \} }$, which is a stronger condition than the usual triangular inequality ${\rm d}_{{\rm ij}}^{{\rm \wedge }} {\rm \; }\le {\rm \; d}_{ik}^{{\rm \wedge }} {\rm \; +\; d}_{{\rm kj}}^{{\rm \wedge }} $ \cite{Rammal1986}. The distance ${\rm d}_{{\rm ij}}^{{\rm \wedge }} $ is called the subdominant ultrametric distance \cite{Benzecri1984,Situngkir2004}. Then, one can construct an HT by using this inequality.

We also use average linkage cluster analysis (ALCA) in order to observe the different clusters of countries according to their geographical location and economic growth more clearly. Since the procedures to obtain ALCA have been presented by Tumminello et al. \cite{Tumminello2010} and Kantar et al. \cite{Kantar2011,Kantar2011a} in detail, we will not explain its construction in here.

\subsection{The measure of link reliability with bootstrap technique}

The bootstrap technique, which was invented by Efron \cite{Efron1979}, has been widely used in phylogenetic analysis since the paper by Felsenstein \cite{Felsenstein1985} as a phylogenetic hierarchical tree evaluation method \cite{Efron1996}. This technique was used to quantify the statistical reliability of hierarchical trees and correlation based networks by Tumminello et al. \cite{Tumminello2007b,Tumminello2007a,Tumminello2010}. Kantar et al. also applied the bootstrap technique to investigate the value of statistical reliability to the links on the hierarchical structures of Turkey's foreign trade \cite{Kantar2011} and major international and Turkish companies \cite{Kantar2011a}.

In order to quantify the statistical reliability of the links of the MST and HT, the bootstrap technique is applied to the data. The number of replicas used in
each periods is 1600. For example, in phylogenetic analysis r = 1000 is usually considered a sufficient number of replicas \cite{Felsenstein1985}. The numbers appearing in Fig. 1 quantify this reliability (bootstrap value) and they represent the fraction of replicas preserving each link in the MST. Recently, this technique used to measure the reliability of the links of MST and HT \cite{Tumminello2007a,Tumminello2007b}. We should also mention that this technique has been well explained in Refs. \cite{Keskin2010a,Kantar2011,Tumminello2007a,Tumminello2007b,Tumminello2010}.

\section{Data}

For the present study we chose 28 countries (the EU27 and Norway) in Europe and used the quarterly debt ratios of these countries for the years 2000-2011. We used data covering the periods 01.01.2000 - 12. 30. 2011, 01.01.2000 - 12. 30. 2004 and 01.01.2005 - 12. 30. 2011, and listed was the countries and their corresponding symbols in Table 1. The quarterly debt ratio downloaded from the Eurostat database, available online ($http://epp.eurostat.ec.europa.eu$).
In the next section, we will construct the MSTs, including the bootstrap values, and the HTs from this data-set and their clustering structures.

\section{Numerical results and discussions}
In this section, the MSTs, including the bootstrap values, and HTs of 28 countries based on the debts of European countries over the period 2000-2011, and two sub-periods, namely 2000-2004 and 2005-2011 were given. The cluster structures by using a clustering linkage procedure were also presented.

The MSTs by using Kruskal's algorithm
\cite{West1996,Kruskal1956,Cormen1990} for the country debts
based on a distance-metric matrix were constructed. The amounts of the links that persist from one node (country) to the other corresponds to the relationship among the countries in Europe. The bootstrap
technique to associate a value of statistical reliability to the
links of the MST were also carried out in which if the values are close to one, the
statistical reliability of the link is very high \cite{Keskin2010a, Tumminello2007a}. The cluster structure of the hierarchical trees were found more clearly within
the ALCA.

\subsection{Hierarchial structures during the 2000-2011 period and subperiods}
Figs. 1a-1c show the MST, applying the method of Mantegna
\cite{Mantegna1999} and Mantegna and Stanley \cite{Mantegna2000}, for
the country debts based on a distance-metric matrix for the periods
of 2000-2011, 2000-2004 and 2005-2011, respectively. Fig. 1a is obtained by using the percentage of Gross Domestic Product (GDP) of the European countries in 2000-2011 period. In Fig. 1a, we observed two different clusters of countries according to their level of debt and economic ties. The first cluster, which has more than 60 $\%$ Maastricht Debt as a percentage of GDP of the country's debt ratio, consists of Germany, United Kingdom, France, Spain, Italy, Greece, Portugal, Austria, Ireland, Belgium, Malta and Cyprus. In this cluster, there is a strong relationship between Austria - Belgium, Spain - France and Belgium - Italy. We can establish this fact from the bootstrap value of the links among the countries, which is equal to 0.90, 0.66 and 0.61 in a scale from zero to one, respectively. In contrast, the bootstrap value of the link between Spain and Germany is very low. In the second cluster are less than 60 $\%$ percent to GDP of the country's debt ratio. The bootstrap value of the link between the Latvia and Romania is equal to 0.98; hence Latvia and Romania are strongly connected with each other. Similar results were also observed in Ref. \cite{Dias2011}. The MST shown in Fig. 1b was obtained for the 2000-2004 sub-period. At the end of this sub-period, some countries, such as Germany, France, Italy, Greece, Austria, Belgium, Malta and Cyprus, are more than 60 $\%$ percent to GDP of the country's debt ratio. In the MST, one cluster is composed of Germany, France, Italy, Austria, Belgium and Cyprus which are more than 60 $\%$ percent to GDP of the country's debt ratio except the Poland and Ireland; hence it is a heterogeneous cluster. The bootstrap value of the link between Austria and Belgium is very high, namely 1.00. The clustering behavior in Fig. 1c is similar to Fig. 1a. Moreover, in Fig. 1c, there is a strong relationship between, Spain - France, Austria - Belgium and Belgium - Ireland, which the bootstrap values of the links are equal to 0.75, 0.70 and 0.60, respectively. Finally we notice that although Ireland and Poland have 29.3 $\%$ and 45.7 $\%$ Debt as a percentage of GDP of the debt ratio at the end of 2004, respectively, that were observed in the heterogeneous cluster of Germany, France, Italy, Austria, Belgium and Cyprus, seen in Fig. 1b. This results indicates that, these two countries are likely to be affected by the crisis, in which Ireland and Poland have 102.3 $\%$ and 56 $\%$ Debt, respectively as a percentage of GDP of the debt ratio at the end of 2011.

The HTs of the subdominant ultrametric space associated with the MSTs
are shown in Fig. 2. Two countries (lines) link when a horizontal
line is drawn between two vertical lines. The height of the
horizontal line indicates the ultrametric distance at which the two
countries are joined. In Fig. 2a, one can observe
two clusters: The first cluster consists of the main countries affected by the crisis that is separated into two sub-groups. The first sub-group contains Italy, Belgium, Austria and Ireland. The distance between Italy and Belgium is the smallest of the sample, indicating the strong relationship
between these two countries. Moreover, the obtained bootstrap
values of the links between these countries is equal to 0.61 which is seen in Fig. 1a. The second sub-group consists of the United Kingdom and Spain, which is weakly connected each other and the bootstrap value is 0.39 seen in Fig. 1a. The second cluster consists of countries of Luxembourg, Latvia, Denmark and Netherlands that were less affected by the crisis in the European area except for the Netherland; hence it is a heterogeneous cluster. The HT presented in Fig. 2b was obtained for the 2000-2004 sub-period. In Fig. 2b, we obtained two clusters without a sub-group. The first cluster contains Italy, Belgium and Austria. The distance between Austria and Belgium is the smallest. The second cluster consists of the United Kingdom, Luxembourg and Finland, which has less than 60 $\%$ Debt as a percentage of GDP of the debt ratio.
Moreover, the behavior of Fig. 2c for the 2005-2011 sub-period is resembling with Fig. 2a, except the second cluster is a heterogeneous cluster because of Netherland and United Kingdom are observed in this cluster. In addition to these two countries, Luxembourg, Romania, Latvia, Lithuania and Denmark are seen in this cluster.

We also obtained more clearly cluster structure by using average linkage cluster analysis (ALCA) for all of the above periods. The obtained results were presented in Figs. 3a-3c. In all periods, we obtained two clusters, namely more than 60 $\%$ Maastricht and less than 60 $\%$ Maastricht. In these figures, we also observe that the Austria, Belgium and Italy are strongly correlated with each other and formed a group in all periods. Moreover, Ireland joined to this group in the 2005-2011 period. Similarly, we found that the Portugal, Sweden and Norway also constituted a different group in all periods.

\section{Summary and conclusion}
We presented the hierarchical structures of countries based on the
debts of European countries by using the concept of the minimal spanning tree (MST), including the bootstrap values, and the hierarchical tree (HT)
for the 2000-2011 period and two sub-periods, namely 2000-2004 and 2005-2011. We obtained the clustered structures of the trees and identified
different clusters of countries according to their level of debts and economic ties. From the topological structure of these
trees, we found that by the debt crisis, the less and most affected Eurozone's economies are formed as a cluster with each other in the
trees. Moreover, similar results were obtained in study using the concept of an MST and HT for nineteen EU countries by Dias \cite{Dias2011}. We performed the bootstrap technique to associate a value of statistical reliability of the links of the MST and HT to obtain information about the statistical reliability of each link of the trees. From the results of
the bootstrap technique, we can see that, in general, the bootstrap
values in the MST are lowly consistent with each other. Furthermore, we found more clearly the cluster structures by using average linkage cluster analysis (ALCA) for the 2000-2011 period and two sub-periods. Finally, we hope
that the present paper will help to give a better understanding of the overall
structure of European debts, and also provide a valuable
platform for theoretical modeling and further analysis.

\begin{acknowledgments}
We are very grateful to Yusuf Kocakaplan for useful discussions.
\end{acknowledgments}

\begin{itemize}
\item [\textbf{Fig. 1(a)}] (Color online) Minimal spanning tree associated to quarterly data of the 28 countries in Europe during the 2000-2011 period.
\item [\textbf{(b)}] (Color online) Same as Fig. 1(a), but for debts as a percentage of GDP during the 2000-2004 period.
\item [\textbf{(c)}] (Color online) Same as Fig. 1(a), but for debts as a percentage of GDP during the 2005-2011 period.

\item [\textbf{Fig. 2(a)}] (Color online) The HT associated with quarterly data of the 28 countries in Europe during the 2000-2011 period.
\item [\textbf{(b)}] (Color online) Same as Fig. 2(a), but for debts as a percentage of GDP during the 2000-2004 period.
\item [\textbf{(c)}] (Color online) Same as Fig. 2(a), but for debts as a percentage of GDP during the 2005-2011 period.

\item [\textbf{Fig. 3(a)}] (Color online) Average linkage cluster analysis. Hierarchical tree associated to a system of the 28 countries in Europe during the 2000-2011 period.
\item [\textbf{(b)}] (Color online) Same as Fig. 3(a), but for debts as a percentage of GDP during the 2000-2004 period.
\item [\textbf{(c)}] (Color online) Same as Fig. 3(a), but for debts as a percentage of GDP during the 2005-2011 period.

\end{itemize}

\begin{center}
\textbf{List of the Table Captions}
\end{center}

\begin{itemize}
\item [\textbf{Table 1}] (Color online) Countries, debts as a percentage of GDP for the 2000-2011 period.
\end{itemize}

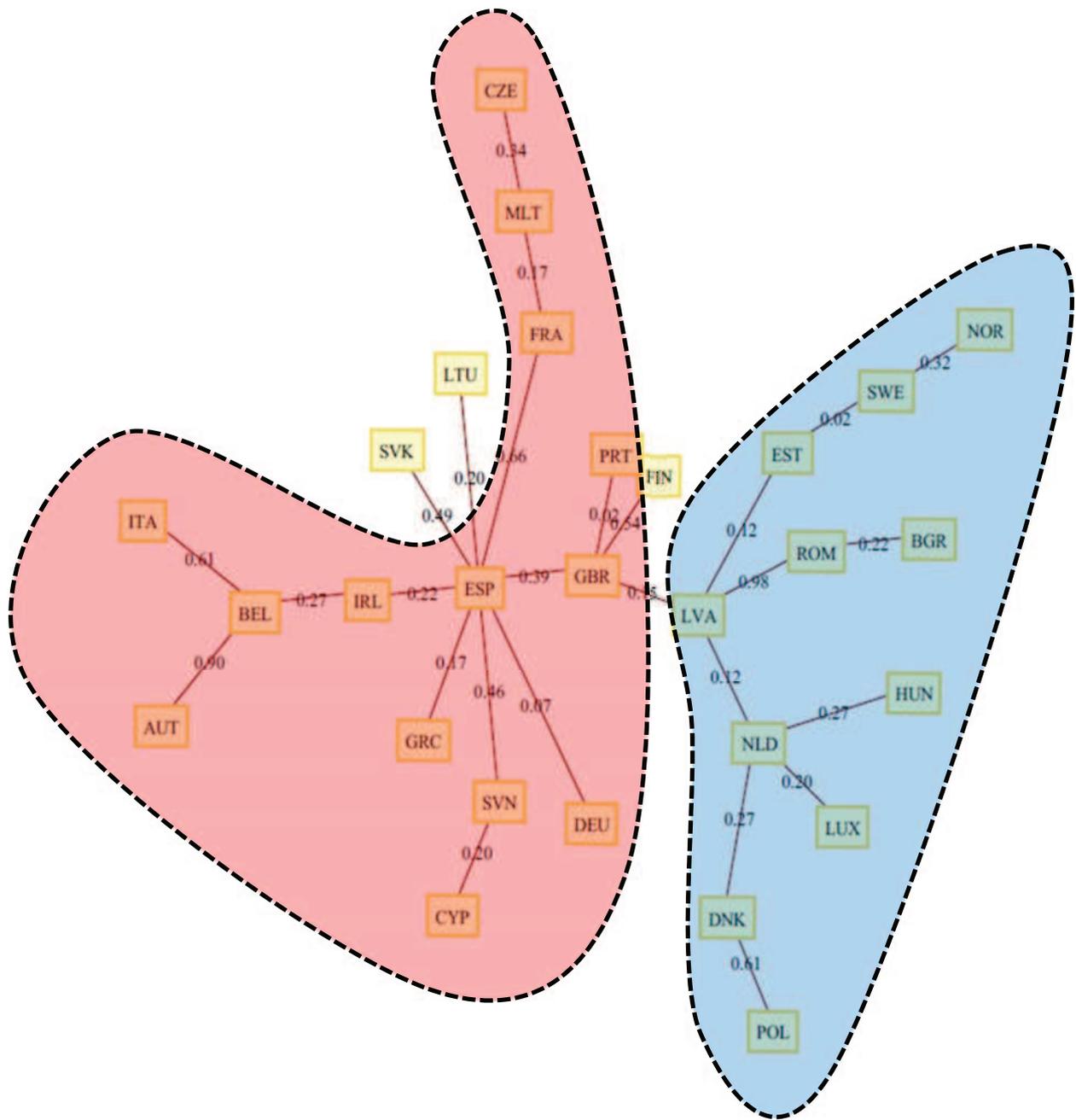
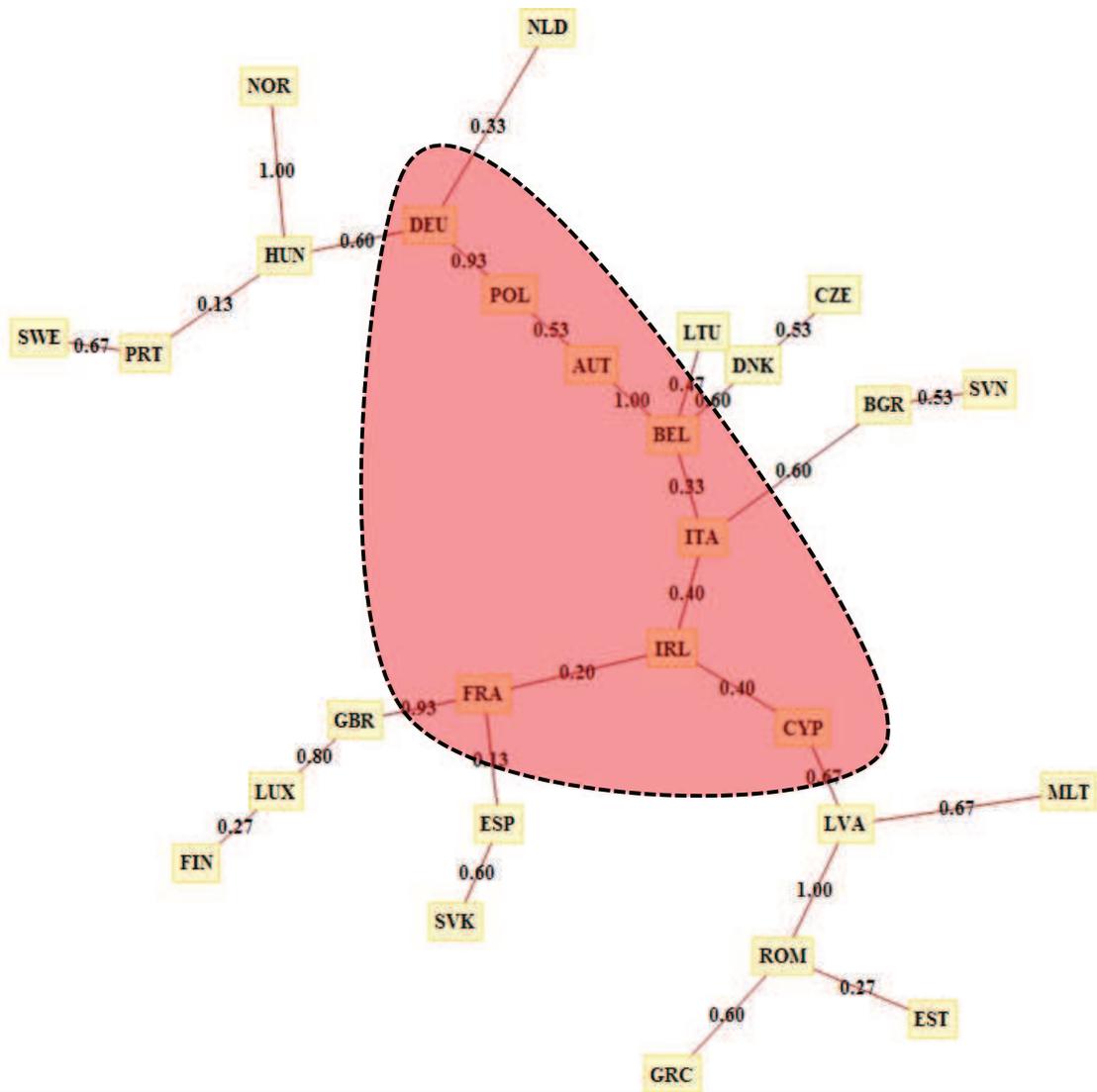
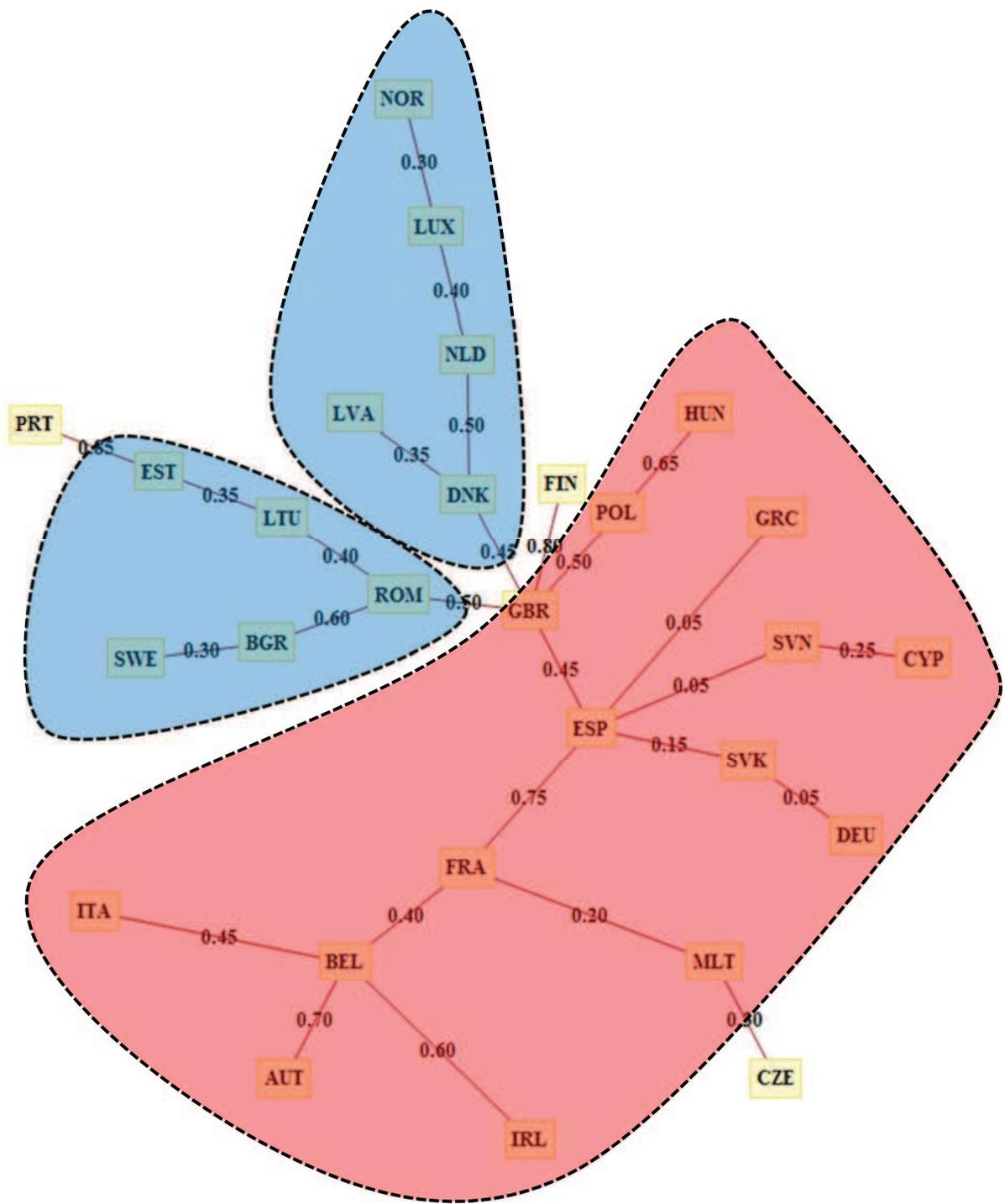
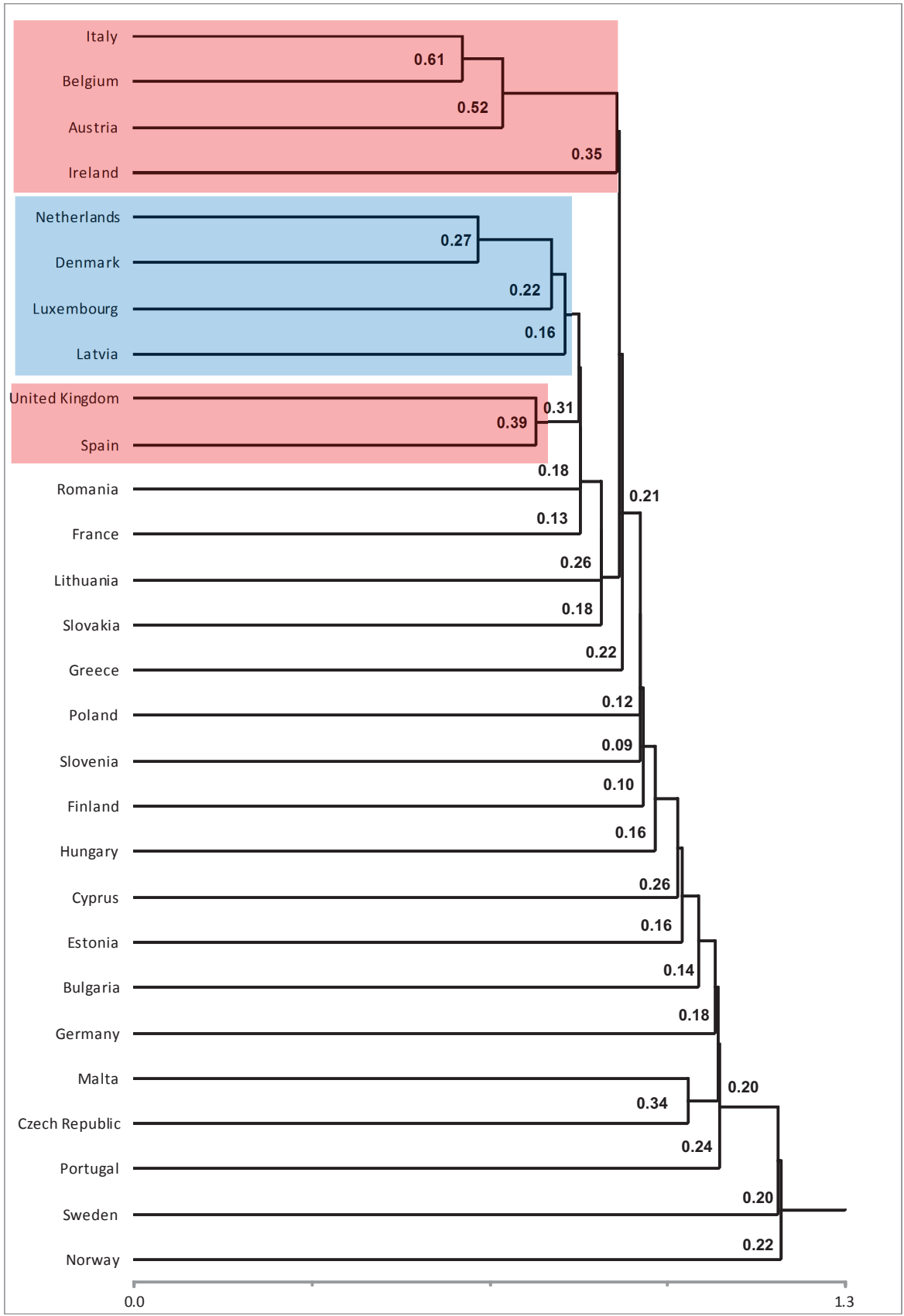
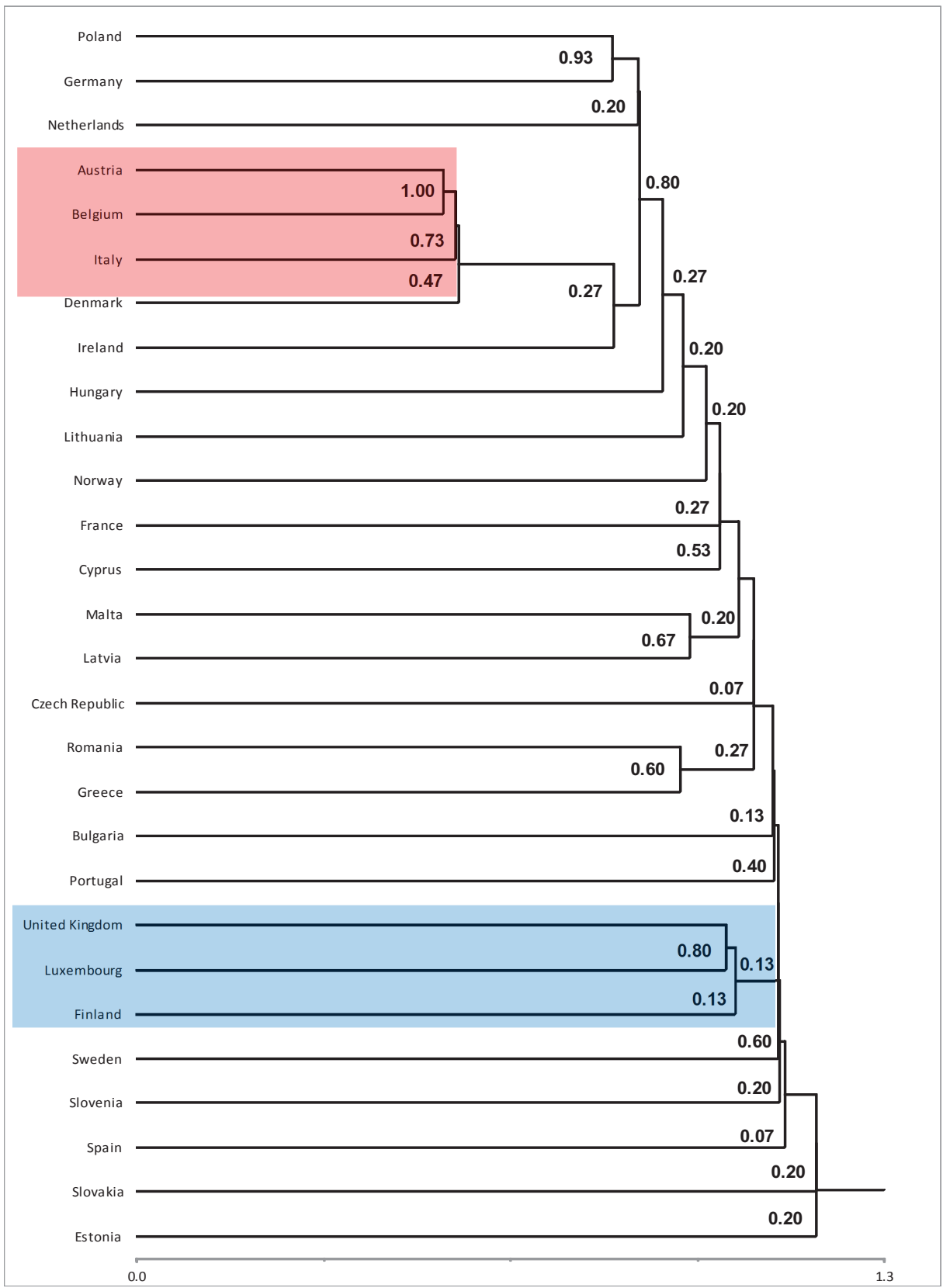
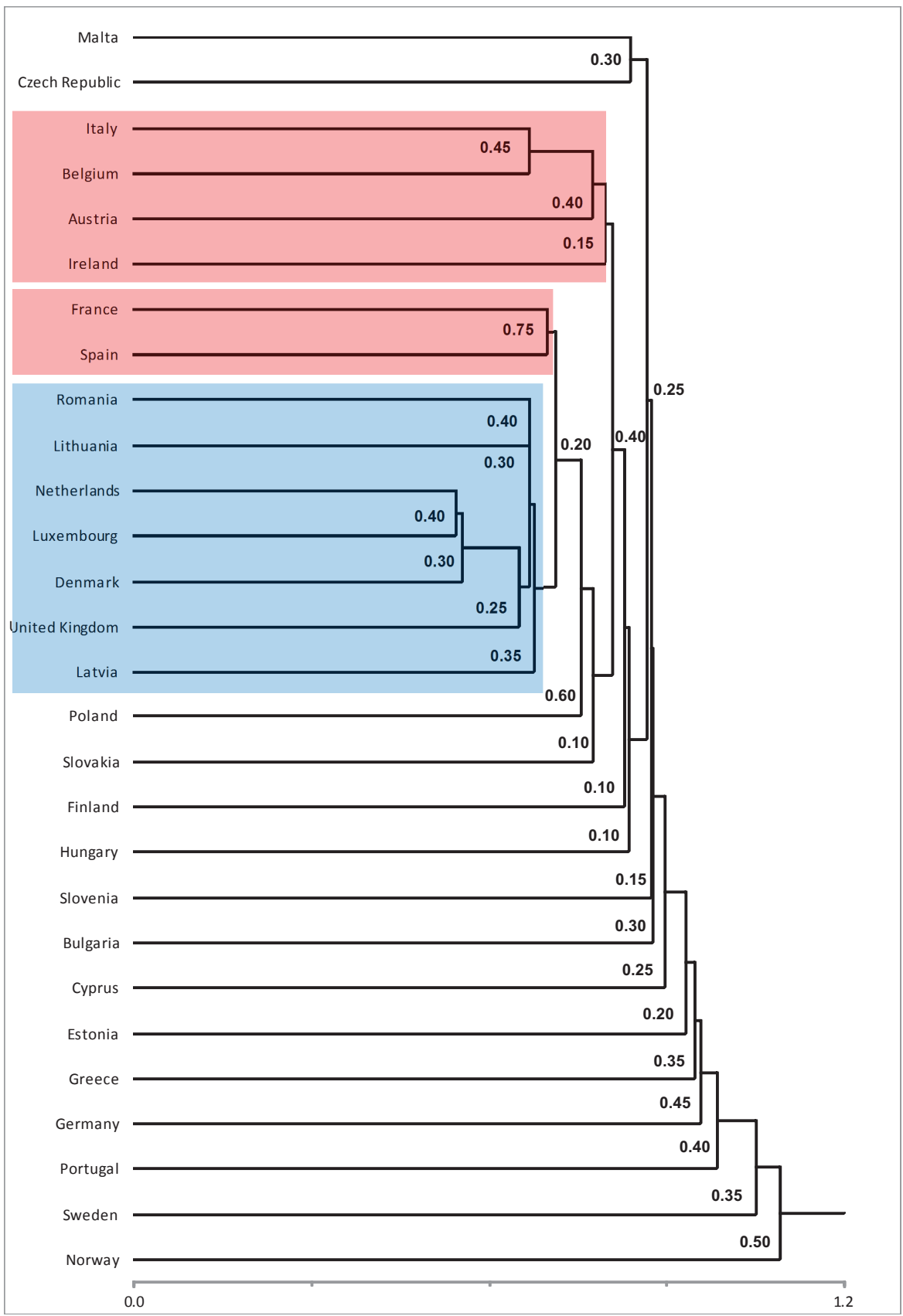
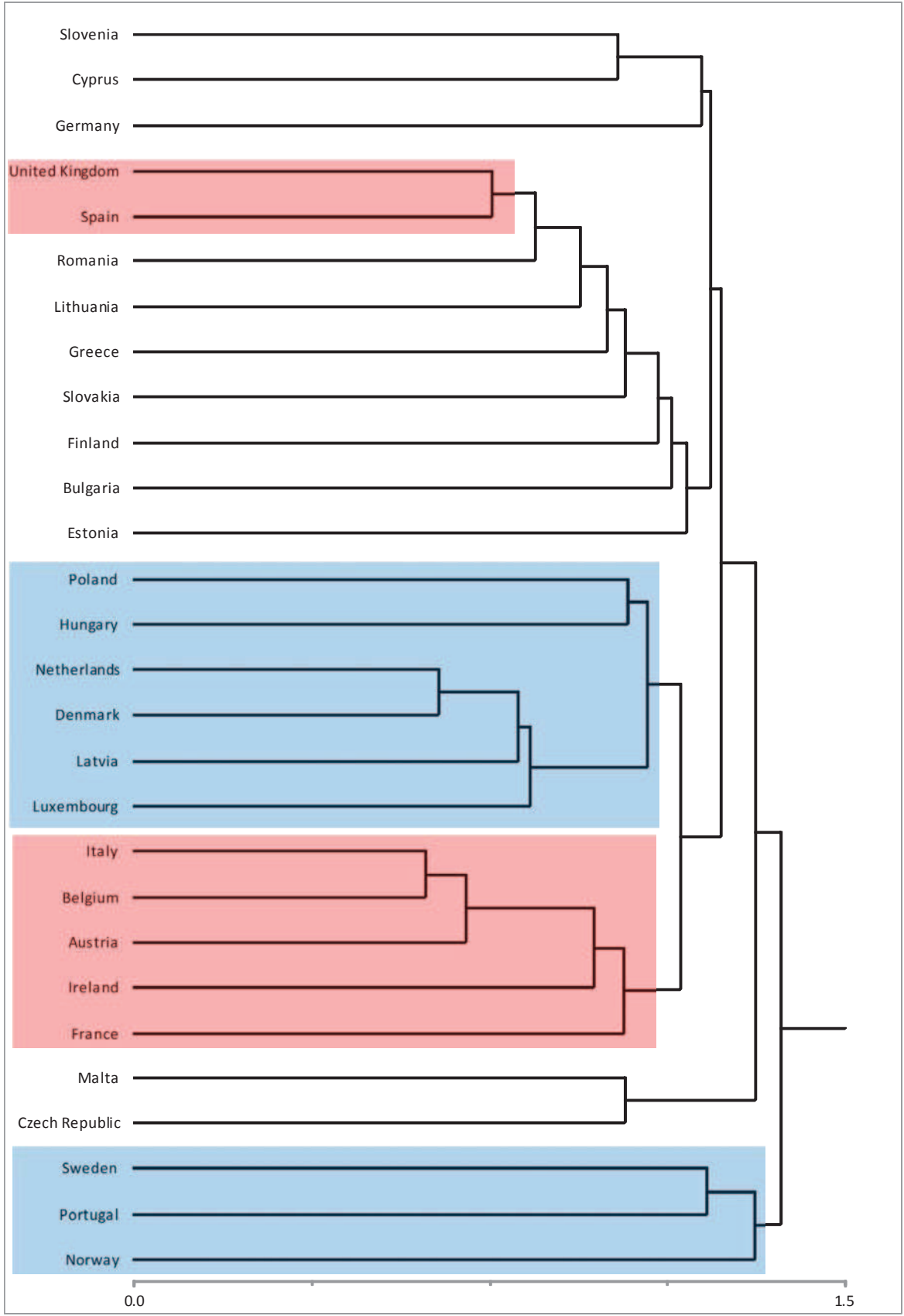
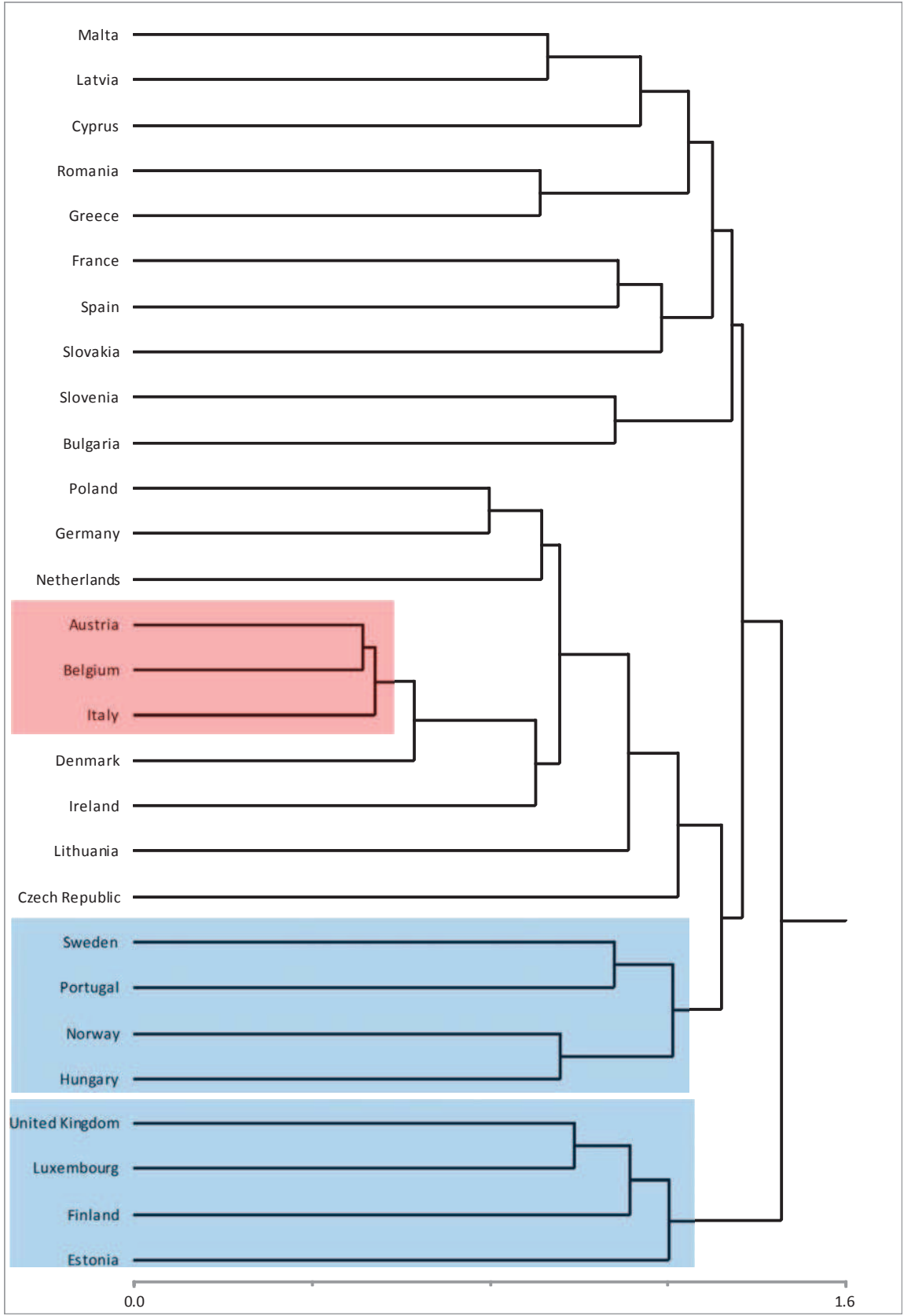
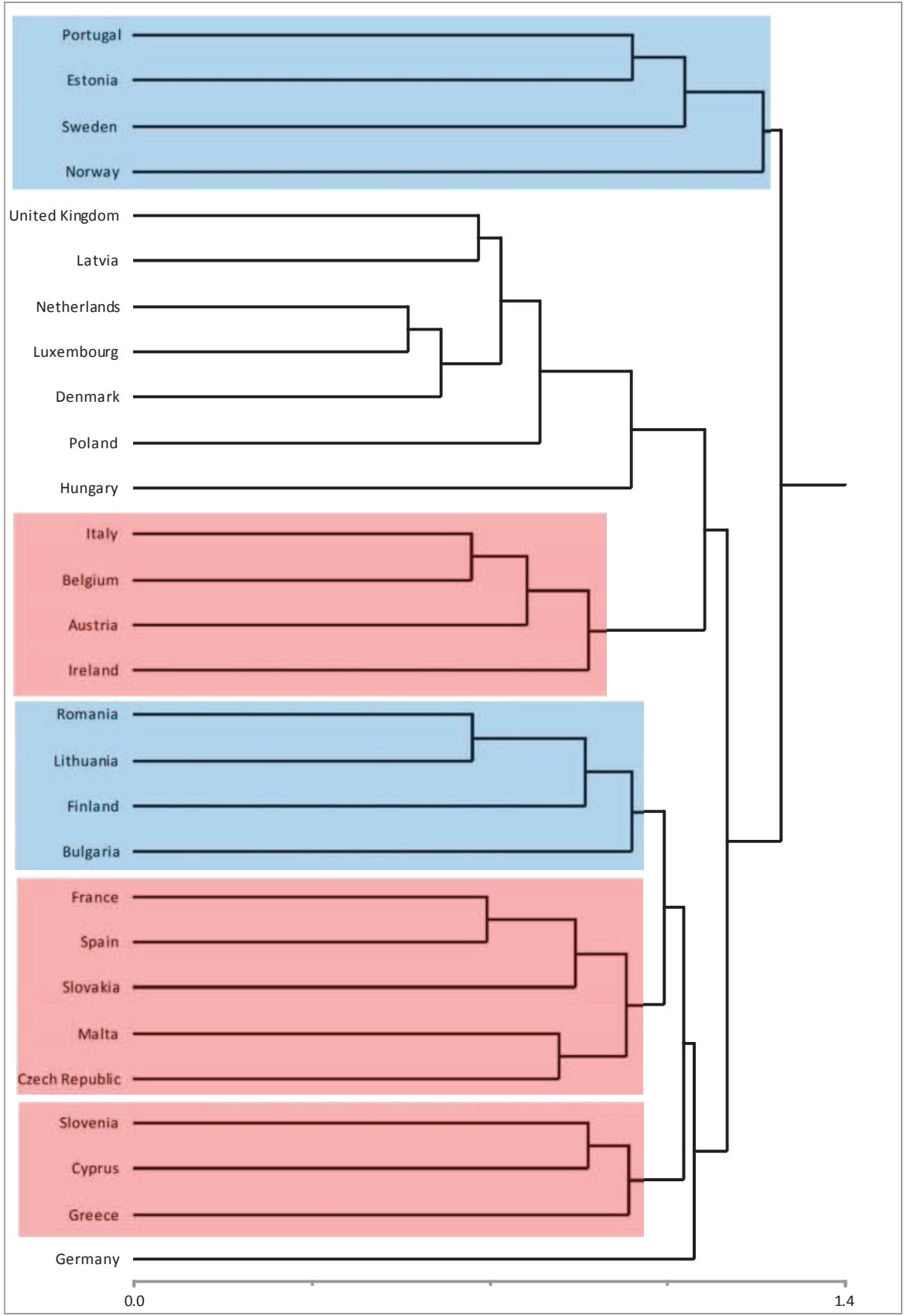
\end{document}